Title: Atom probe composition and *in situ* electronic structure of epitaxial quantum dot ensembles

Authors: Christopher Natale, Ethan Diak, Ray LaPierre, Ryan B. Lewis*

Department of Engineering Physics, McMaster University, L8S 4L7 Hamilton, Canada

*Email: rlewis@mcmaster.ca

Abstract

Dense arrays of semiconductor quantum dots are currently employed in highly efficient quantum dot lasers for data communications and other applications. Traditionally, the electronic properties of such quantum nanostructures have been treated as isolated objects, with the degree of hybridization between neighboring quantum dots and the wetting layer left unexplored. Here, we use atom probe tomography and transmission electron microscopy to uncover the three-dimensional composition profile of a high-density ensemble of epitaxial InAs/GaAs quantum dots. The sub-nanometer compositional data is used to construct the 3D local band structure and simulate the electronic eigenstates within the dense quantum dot ensemble using finite element method. This *in situ* electronic simulation reveals a high degree of hybridization between neighboring quantum dots and the wetting layer, in stark contrast to the usual picture of isolated quantum nanostructures. The simulated transition energies are compared with low temperature photoluminescence. This work has important applications for quantum dot laser design and paves the way to engineering ensemble effects in quantum dot lasers and other quantum nanostructures.

Introduction

Epitaxial quantum dot (QD) lasers have many advantages over conventional quantum well lasers, including lower threshold currents, better temperature stability and higher data rates [1]. Importantly, carrier localization in QDs reduces the sensitivity to lattice defects, making QD lasers promising for heteroepitaxial integration with Si photonic integrated circuits [2]. Charge carriers in QDs are confined in three dimensions (3D), resulting in discrete energy eigenstates, which dictate exciton recombination energies as well as carrier dynamics and device stability [3]. The emission spectrum can be tuned by controlling the QD size, composition and the surrounding matrix (dot-in-well structure [4]), thereby altering the energy eigenstates within the ensemble.

Coupled electrostatic QD systems have been explored for quantum information and computing technologies, where quantum gates are formed with the application of external fields [5,6]. Quantum dot coupling has also been explored in vertically-aligned self-assembled InAs/GaAs QDs, for which entangled photon emission has been demonstrated [7]. Coupling requires coherent tunnelling, forming wavefunctions that are delocalized across multiple QDs [8]. Symmetric and anti-symmetric molecular wavefunctions analogous to molecular bonds have been demonstrated in a vertically-stacked InAs/GaAs QD pair [8]. Similar behavior has also been observed in core-shell colloidal QD systems [9]. The interaction between QDs has been modelled using a linear combination of QD orbitals for a large electronically coupled chain, whereby the carriers are paired through occupying a ground state within one QD and an excited state within the adjacent QD [10]. One application of QD tunneling is the tunnelling injection QD laser, which has improved temperature stability and performance over traditional QD lasers [11]. In general, engineering tunneling and hybridization in QDs requires extremely accurate control of the nanostructure properties.

Viable QD lasers require high planar QD densities—typically $\gg 10^{10}$ QDs/cm$^2$—to achieve sufficient gain required for lasing. At such high QD density, the close proximity of neighboring QDs is expected to lead to coupling and hybridization between QDs [12,13], for which the impact on laser device performance has not been thoroughly explored. Atom probe tomography (APT) can provide a 3D composition reconstruction of QDs in their surrounding matrix; however, APT is limited to lateral dimensions of a few tens of nanometers, necessitating an extremely high planar QD density to simultaneously measure multiple QDs in a sample. While single QD APT data has been used to simulate isolated QD eigenstates [14–19], to our knowledge, in-plane QD coupling within a single Stranski–Krastanov (SK) QD layer has not been reported. The finite element method (FEM) has been invoked to solve energy states for the carriers through the Schrödinger equation, as it has been proven to accurately model the electronic and optical properties for a variety of complex QD systems [20–27].

In this work, we use APT to uncover the sub-nanometer 3D composition profile of high density InAs SK QDs, and use this to simulate electron and hole eigenstates *in situ* with FEM. We demonstrate complex in-plane hybridized multiple QD eigenstates. This work opens the door to engineering hybridized eigenstates in other complex nanostructures, such as a coupled QD-ring pairs [28,29] and other QD material systems.

Methods

Samples were grown on GaAs(001) substrates by gas-source molecular beam epitaxy (MBE) using an SVTA-MBE35 system. Indium and gallium effusion cells were used for the group III atoms, and thermally-cracked phosphine and arsine were used to produce dimers for the group Vs. Before growth, the substrate was thermally annealed under a hydrogen plasma for 10 minutes at 440 °C while flowing 2 sccm of cracked $AsH_3$, to desorb the native oxide. The substrate temperature was subsequently lowered to the growth temperature of 400 °C. The sample contained three buried InAs QD layers and a single surface layer of InAs QDs with the following layer structure: GaAs(001) substrate / GaAs (100 nm) buffer / $In_{0.5}Ga_{0.5}P$ (25 nm) / GaAs (100 nm) / InAs [1.7 monolayers (ML)] / GaAs (20 nm) / InAs (1.7 ML) / GaAs (20 nm) / InAs (1.7 ML) / GaAs (100 nm) / $In_{0.5}Ga_{0.5}P$ (25 nm) / GaAs (100 nm) / InAs (1.7 ML). All $In_{0.5}Ga_{0.5}P$ and GaAs layers were grown at a rate of 0.28 nm/s and a V/III flux ratio of 2. The SK InAs QD layers were grown at a rate of 0.1 ML/s by opening the indium source for 1.8 seconds, closing the shutter for 6.7 seconds, and repeating this process twice, thereby having each open-close event depositing half the target number of monolayers.

The sample was imaged by transmission electron microscopy (TEM) in high angle annular dark field (HAADF) mode on a Thermo Scientific Talos 200X TEM. Photoluminescence spectra were collected at 7.7 K using argon ion laser excitation at 488 nm. APT specimens were prepared using a plasma focused ion beam (pFIB) to shape the specimen into a fine tipped needle. The pFIB uses Xe ions to avoid altering the chemical composition of the sample. APT experiments were carried out in a local electrode atom probe (LEAP) 5000 XS system, which is a straight-flight-path APT with spatial resolution $\Delta z \approx 0.1\text{--}0.3$ nm and $\Delta x, \Delta y \approx 0.3\text{--}0.5$ nm, where z corresponds to the substrate normal direction, <001>. For a detailed description of the APT experimental process, the reader is referred to ref. [30]. The APT specimen was held at a temperature of 40 K while a 355 nm laser with pulse rate of 250 kHz and pulse energy of 0.3 pJ was used for ablation. The detection rate was held constant at 0.5%, gathering over 23 million total detection events. Matlab was used for processing the entire dataset to generate a FEM model using COMSOL which solves the 3D time-independent Schrödinger equation.

FEM decomposes any arbitrary subset of space into smaller volumetric entities that are interconnected at nodal points, such as boundaries, faces, edges, and vertices. Each of the finite elements has an associated partial differential equation governed by the boundary value problem assigned to it. These are combined into a larger system of equations that converge to a solution by minimizing the overall error of the eigenvalue problem by assessing the local stability around an equilibrium point through linearization.

The 3D time-independent Schrödinger equation for the electrons and heavy holes is

$$\left[\frac{-\hbar^2}{2m_{e/hh}(x,y,z)}\nabla^2 + V_{e/hh}(x,y,z)\right]\Psi_{e/hh}(x,y,z) = E_{e/hh}\Psi_{e/hh}(x,y,z) \quad (1)$$

where the effective mass of electrons/heavy holes, $m_{e/hh}$, has been explicitly made a function of the position. The light holes have been neglected due to the much larger density of states for the heavy hole band. The APT data set is used to assign the local potential energy, $V_{e/hh}(x,y,z)$, and effective mass throughout a high density InAs/GaAs(001) quantum dot layer.

Results and Discussion

Fig. 1(a) shows a cross-sectional HAADF TEM of the entire heterostructure, illustrating the three QD layers, between the two InGaP cladding barriers with GaAs spacers. In Appendix A, a high-resolution TEM image is presented of a single QD with radial height ~2.5 nm, radial width ~6 nm and QD core composition $In_{0.31}Ga_{0.69}As$. Fig. 1(b) displays a scanning electron microscope (SEM) image of an APT specimen, prepared from a separate region of the sample. The needle is aligned such that the axial direction is approximately along the <001> substrate normal. Fig. 1(c) shows a 3D APT scatter plot of In and Ga counts in the region surrounding the three QD layers (see Appendix B for the entire APT data set). The InAs QD layers (white) appear slanted, presumably due to a slight misalignment during pFIB preparation of the APT specimen from the <001> growth axis. A 10 nm thick in-plane APT slice of the middle QD layer is presented in Fig. 1(d), which has been rotated such that the viewing axis is down the growth axis. The QDs are visible as regions of denser indium counts, consistent with QDs embedded within a thin InGaAs wetting layer. Atomic force microscopy (AFM) measurements on the InAs surface QDs from this sample—grown at the same conditions as the buried QD layers—reveal a QD density of $1.8 \times 10^{11}$ dots/cm². The APT measurements indicate a local QD density (within the middle layer) of ~$7 \times 10^{11}$ dots/cm²—possibly larger than AFM as APT can measure very small dots localized in the wetting layer, or due to post-growth ripening of the surface QDs. The surface dots exbibit an AFM height of 1.2 ± 0.7 nm and radial width of 7.6 ± 2.4 nm. We note that the AFM height is less than that observed by TEM, due to the wetting layer thickness not being included in the AFM height. Overall, these results indicate a dense array of small QDs, owing to the low InAs growth temperature.

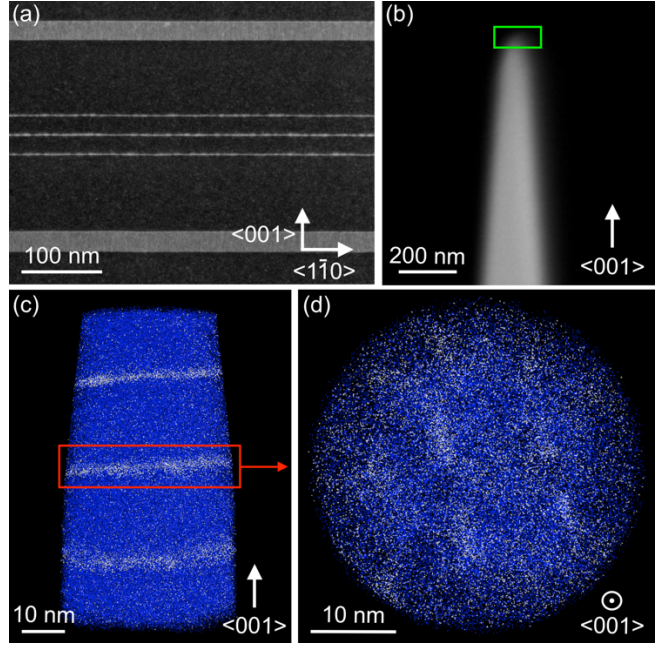

FIG. 1. (a) Cross-sectional HAADF TEM image along the <110> zone-axis. (b) SEM image of the APT sample. The green box roughly corresponds to the heterostructure. (c) APT scatter data of Ga ions (blue) and In ions (white) showing the three QD layers. (d) In-plane APT scatter data of the middle QD layer (10 nm slice thickness).

To assign a 3D composition corresponding to the In and Ga APT scatter data, the sample volume of interest (red box shown in Fig. 1(c)) was decomposed into a grid. For each 3D rectangular grid element (voxel) an effective indium concentration was computed based on the fraction of atoms contained within the voxel, as well as neighboring voxels according to a defined searching algorithm. Here, "neighbor" refers to any voxel touching the central voxel's corners, faces, or edges. Each voxel must have an assigned composition, since the entire sample volume must be assigned an associated bandstructure. Our algorithm initially searches the central voxel, as well as the nearest neighboring voxels, for In and Ga atoms. If at least one atom is found, then a composition is assigned to the central voxel. If no atoms are found, the algorithm subsequently increases the search radius to next-nearest neighbors, etc., until at least one Ga/In atom is detected, so that a composition can be assigned to the central voxel. Forcing the algorithm to search first nearest neighbors regardless of whether the central voxel contains any atoms is implemented as a form of boxcar smoothing, also known as delocalized smoothing [31]. This approach enables arbitrarily small voxel sizes to be chosen, while still assigning a composition to the entire simulation volume. Fig. 2 depicts the probability that a voxel requires a search radius of n=0, n=1, n=2 and n=3 neighbors to find at least one group III atom, plotted as a function of the voxel volume. As the voxel volume decreases, the probability that more neighbor searching is required increases. We note that both decreasing the voxel size and increasing the required neighbor searching increase the required computational resources. To generate the probability distribution in Fig. 2, a 28×28×10 nm³ region of the APT data set encompassing the middle QD layer was utilized (shown in Appendix C). The method for selecting voxel dimensions is further discussed in Appendix C.

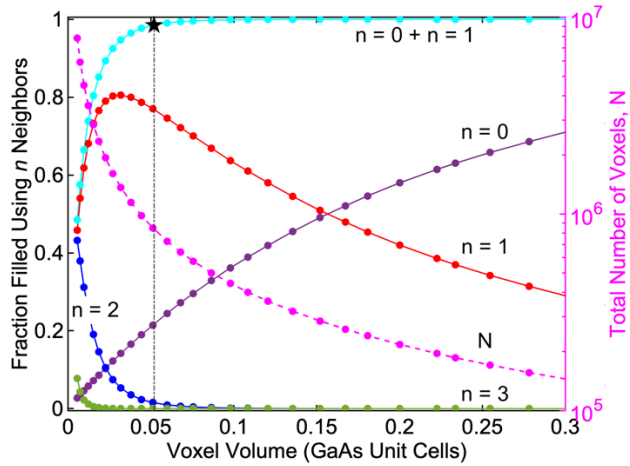

FIG. 2. Probability that exactly *n* nearest neighbors is required to detect a group III atom, plotted as a function of voxel volume. The vertical line at ~$0.05 a_0^3$ corresponds to the selected voxel size. The total number of voxels in the sample volume is plotted on the right-hand scale.

For subsequent analysis, a voxel volume of ~$0.05 a_0^3$ was used—vertical line indicated in Fig. 2. Defining the model spatial resolution as the distance along the voxel body diagonal, this corresponds to an average spatial resolution of $1.1 \pm 0.1$ nm (see Appendix C for details). Due to the very small size of the QDs, it is important to choose a voxel size which maximizes the spatial resolution to accurately model the electronic properties of the QD ensemble through FEM.

However, an artifact inherent to APT is that the central component of the data cone is denser than the outer region, due to signal compression (demonstrated in Appendix C) [30,32]. The chosen voxel size results in a composition being assigned to 98.5% of all voxels with a single neighbor search (c.f., black star in Fig. 2). Appendix C discusses the variance in spatial resolution and the impact of imposing a minimum neighbor search constraint. The spatial resolution of the model was larger than the APT instrument resolution by ~0.5 nm. We note that the model spatial resolution converges to the APT instrument resolution as the voxel size approaches zero (Appendix C). This is analogous to a Voronoi tessellation, as each sub-volume in the sample space converges to the nearest data point, giving the best possible spatial resolution for a model, but resulting in a digital alloy (voxels composed of pure InAs or GaAs).

Fig. 3(a)–(c) depicts the electron potential energy landscape obtained using the composition assigned by the APT model. The confining potentials are derived using the bandgap and offsets of strained $In_xGa_{1-x}As$ (see Appendix D for details). Fig. 3 exemplifies the *in situ* chaotic nature of the band profile within the QD plane, with voxels ranging in composition from binary InAs/GaAs to ternary alloys of $In_xGa_{1-x}As$. We note the tradeoff between maximizing spatial resolution—resulting in a digital alloy of InAs and GaAs voxels—and representing the average composition of a region. Importantly, while composition alloy fluctuations are significant in our data set, their spatial extent is small in comparison to the exciton Bohr radius depicted in Fig. 3(c) (calculated in Appendix D). Correspondingly, it is expected that these fluctuations will be averaged over while solving for the electron eigenstates. Furthermore, we note that the Bohr radius spans multiple QDs, due to the high QD density.

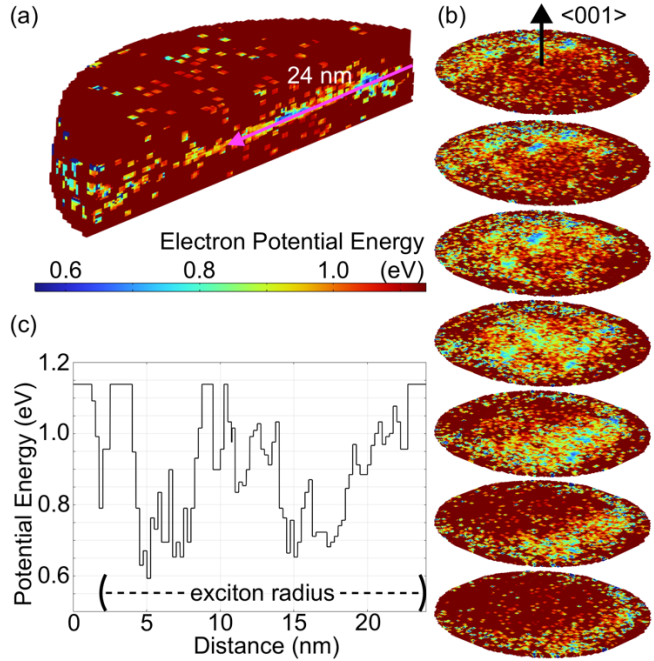

FIG. 3. Electron potential energy, $V_e$, for the middle QD layer plotted in 3D, 2D and 1D. (a) 3D cut through the QD layer. (b) 2D in-plane slices of the QD layer each with 0.3 nm thickness and a diameter of ~42 nm. Due to the slight misalignment of the z-axis from the substrate normal, the slices cut through the QD plane at a small angle. (c) 1D linescan within the QD plane, corresponding to the 24 nm magenta arrow in (a). The exciton radius for pure InAs (~22 nm) is shown for comparison to the dot spacing.

Fig. 4(a) depicts a 3-dimensional map of the local regions with high indium density (labelled A-J) within the middle QD layer. Labels are solely intended as an aid to the reader when visualizing the QD eigenstates. This graph has been created by only plotting voxels containing >50% indium. Furthermore, statistical noise was reduced from the image by establishing an upper limit of 99.8%, since few pure InAs voxels contribute to the QDs, tending to be more prevalent in regions of lower atomic count density near the edges. The orientation of Fig. 4(a) is the same as presented in Figs. 3(a), 3(b) and 4(b) to provide a direct comparison. The linescan depicted in Fig. 3(c) intersects the regions B and D in Fig. 4(a).

Fig. 4(b) presents the probability density for the 8 lowest energy electron and heavy hole eigenstates calculated for the QD ensemble in Fig. 4(a). The eigenstates—corresponding to the lowest energy electron and heavy hole states in this QD layer—are not localized to individual QDs, but instead have considerable probability densities covering two or more QDs. This indicates a high degree of in-plane hybridization between QDs for this sample of high density QDs. Comparing the electron and heavy hole eigenstates, a noticeable feature is that the heavy holes are more localized to the cores than the electrons, due to their heavier effective mass. Interestingly, the order of the electron eigenstates (arranged by eigenenergy) does not match that for the heavy hole eigenstates. The wavefunctions for both the electrons and heavy holes span multiple QDs. There exists a complex arrangement of states and orbital patterns, which would not exist in isolated quantum dot systems. However, we expect that as the density of quantum dots decreases, the interactions between QDs would decrease, with eigenstates becoming more localized to individual QDs.

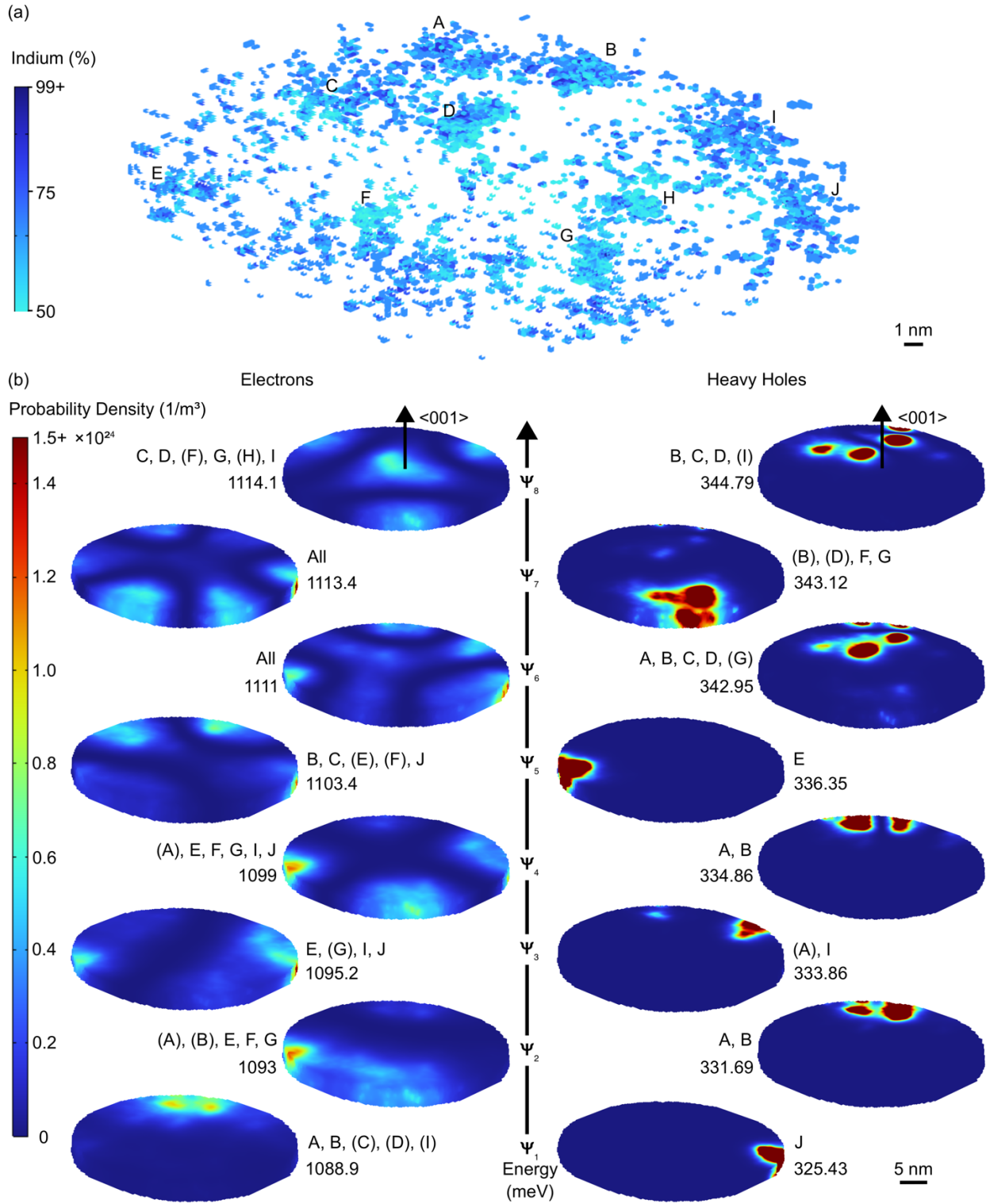

FIG. 4. (a) 3D indium concentration map. Locally dense regions of indium are labelled A-J. (b) Probability density, $|\Psi|^2$, of the 8 lowest energy electron and heavy hole eigenstates

(arranged by energy). The regions from (a) with considerable involvement are listed next to each state. The regions with minor involvement are noted in parentheses.

The simulated eigenstates exhibit both bonding and anti-bonding orbitals between neighboring QDs. In bonding orbitals, $|\Psi|^2$ is nonzero everywhere between the bonded QDs. In anti-bonding orbitals, $|\Psi|^2$ becomes zero between the anti-bonded QDs, corresponding to a node. As an example, $\Psi_{2hh}$ and $\Psi_{4hh}$ demonstrate two states where regions A and B are bonded and anti-bonded respectively. Two higher energy states (involving more QDs) are $\Psi_{6hh}$ and $\Psi_{8hh}$. Together, they show that regions B, C, and D bond in $\Psi_{6hh}$, whereas in $\Psi_{8hh}$, regions C and D remain bonded, while B is anti-bonded to the others. As expected, states where the majority of QDs are bonded have lower energies than states containing QDs that are anti-bonded. We note that state $\Psi_{5e}$ makes use of two distant QDs (B and C) separated by ~15 nm. Even more perplexing, is that there is a QD (A) in between, for which there is negligible probability of finding the electron, as a node line is formed which intersects A.

Another intriguing feature is that $\Psi_{6e}$ and $\Psi_{7e}$ have close energy and similar orbital patterns, connecting many QDs which extend beyond the Bohr radius of the exciton. This potentially shows the emergence of wetting layer states, where the electrons are delocalized among many QDs. We note that linking QDs at greater distances is an important requirement for devices which use QDs to transfer information, such as quantum repeaters [33].

A low temperature (7.7 K) photoluminescence spectrum taken from the grown sample is shown in Fig. 5. The dataset has been boxcar smoothed with a 4 nm box size to remove high-frequency noise from the collected spectrum. The PL peak is located at ~940 nm with a FWHM of 35 nm. All possible simulated transitions for the solved eigenstates are shown in Fig. 5, ranging in energy from ~851—877 nm. We note that Coulombic forces—not taken into account in our simulation due to the computational complexity of the model already—would reduce the simulated recombination energies, shifting the transitions in the direction of the PL emission [34–36]. Furthermore, it is likely that carriers diffuse a much larger distance in the wetting layer than the APT sample area, allowing carriers to find the largest quantum dots before recombining, thus red-shifting the emission.

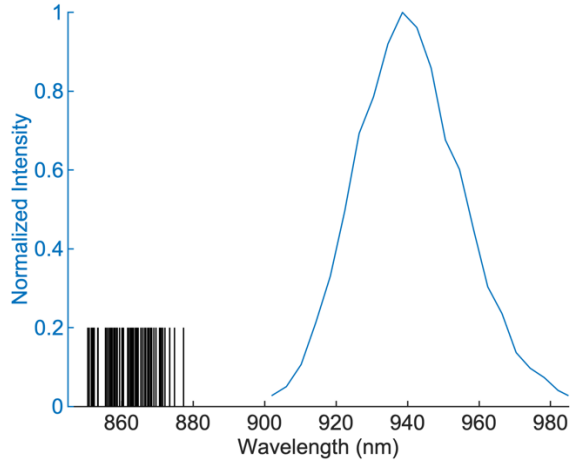

FIG. 5. PL emission at 7.7 K (blue) and all possible simulated transitions (black lines) from the calculated eigenstates.

The simulated eigenstates illustrate how interaction between neighboring QDs can impact the energy, distribution and symmetry of the eigenstates, which could have important implications for the optical properties of the QD ensemble and related devices. The ability to engineer eigenstates through QD hybridization (e.g., by forming QD chains on offcut substrates) could enable novel properties in optoelectronic devices. These findings demonstrate that the interaction between neighboring QDs is highly relevant in dense QD ensembles. Although higher QD densities allow for more potential emitters, it increases the probability of hybridization. Therefore, when designing QD lasers, the exciton Bohr radius and quantum dot density should be considered together since hybridization can cause broadening of the emission spectrum by creating more complex arrangements of orbital patterns. This effect could be beneficial if a broader gain spectrum is desired. For typical 1.3 μm emitting InAs/GaAs(001) QD lasers, the QD density is lower than for the present study, stemming from the need for larger dots. Ref. [37] demonstrated a very high density ($5.9 \times 10^{10}$ dots/cm$^2$) of QDs emitting near 1.3 μm—although this value is much lower than the $7 \times 10^{11}$ dots/cm$^2$ presented here. However, for $5.9 \times 10^{10}$ dots/cm$^2$, we note that the entire 42 nm diameter APT in-plane area (c.f. Figs. 3 and 4) would contain less than a single dot on average.

Conclusion

Combining sub-nanometer-resolution 3D atom probe tomography composition with numerical finite element method (FEM) allowed for *in situ* calculation of the electronic eigenstates of quantum nanostructures in their local environment. We demonstrated a method to create 3D composition profiles with optimized spatial resolution for InAs/GaAs quantum dot layers using a neighbor searching algorithm to avoid vacant volume elements and balance the competition between spatial resolution and digitization of the alloy. The *in-situ* FEM electronic calculations reveal a high degree of hybridization between neighboring QDs for the investigated high-density QD layer. This work has implications for engineering the optical and electronic properties of quantum nanostructures. Furthermore, this *in situ* modeling technique—covering a sample volume of ~$10^4$ nm$^3$ with over 1.3 million volume elements—can be applied to explore ensemble effects in QDs and other systems of complex 3D nanostructures which are otherwise difficult to explore.

Acknowledgments

The authors are grateful to Gabriel Arcuri for APT support, Natalie Hamada for TEM support, and Shahram Ghanad-Tavakoli for MBE support. We acknowledge support from McMaster University's Centre for Emerging Device Technologies and the Canadian Centre for Electron Microscopy. We are grateful for financial support from the Natural Sciences and Engineering Research Council of Canada under project [RGPIN-2020-05721] and the National Research Council's High Throughput Secure Networks Program.

APPENDIX A: Z-CONTRAST HAADF

Fig. 6(a) shows a high-resolution HAADF TEM image of a single exemplary QD, which was used to estimate the composition of the core through Z-contrast analysis. This core is in the middle layer, the same layer analyzed in the main text. Intensity linescans of this QD along the <001> and <1$\bar{1}$0> directions are shown in Figs. 6(b) and (c), respectively.

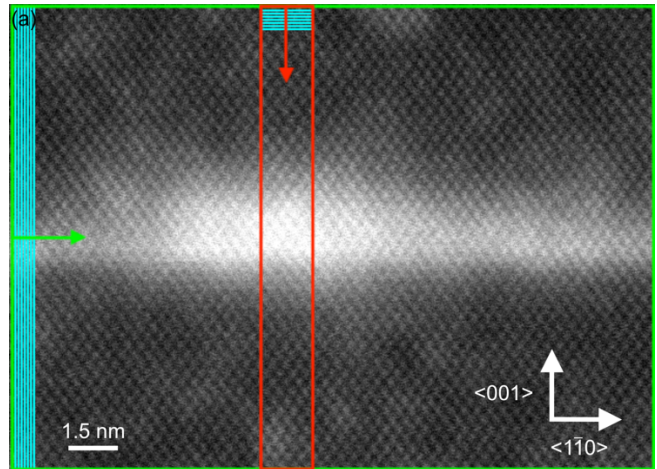
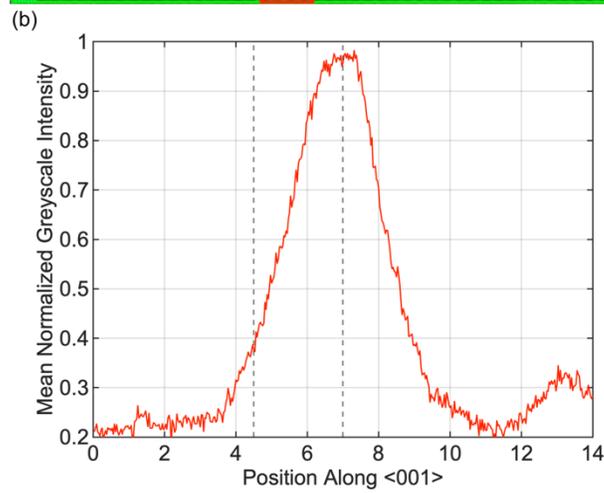
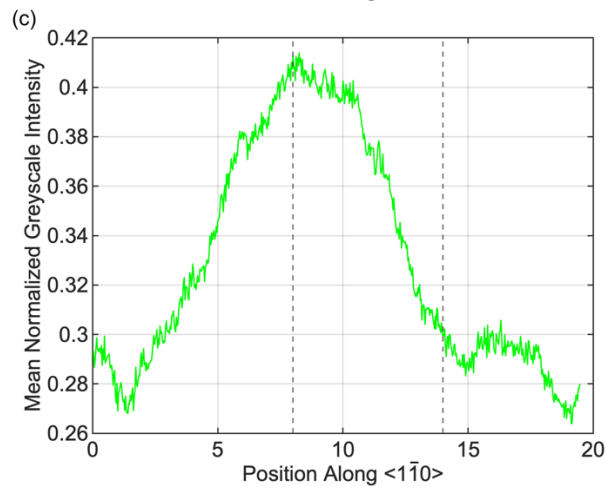

FIG. 6. (a) Cross-sectional HAADF TEM image along the <110> zone-axis of the middle QD layer. The red/green boxes indicate the regions that were used to generate the linescans in (b) and (c), respectively. The arrow indicates the direction of the linescan. Light blue lines are depicted to represent the row/column of pixels being averaged for each data point. The normalized greyscale intensity profile for the QD in (a) is shown for the <001> growth axis (b)

and for the <1$\bar{1}$0> axis (c). The radial height in (b) and radial width in (c), depicted as the distance between dashed lines, has been calculated using 2 standard deviations derived from the FWHM.

The intensity linescans in Figs. 6(b) and (c) have been calculated by converting Fig. 6(a) to an 8-bit greyscale map, with each pixel ranging from 0 (black) to 255 (white). The average of each column/row of pixels (light blue lines in Fig. 6(a)) has been used to create the intensity profile, which has been normalized to the 8-bit maximum (255). This allows the peak intensity, representing the QD core, to be compared to the surrounding background matrix to extract the width, height and core concentration. Figs. 6(b) and (c) use this method to create an intensity linescan along the growth axis and within the wetting layer plane, respectively. Fig. 6(b) shows that the QD has a radial height (along <001>) of ~2.5 nm and Fig. 6(c) demonstrates a radial width (along <1$\bar{1}$0>) of ~6 nm.

The core of the quantum dot in Fig. 6(a) appears brightest, indicating higher local In concentration. The maximum greyscale intensity is $I_{max} \approx 0.97$ depicted in Fig. 6(b). $I_{max}$ has been normalized to the 8-bit greyscale maximum value of 255 (pure white). The intensity for the GaAs regions is $I_{background} \approx 0.2$. Subtracting this yields a peak intensity for the QD core of $I_{core} \approx 0.77$, which is the maximum contrast from the background, $C$.

$$C = I_{core} \tag{A1}$$

The contrast is directly related to the concentration, $c_{In}$, of the alloying element [38].

$$C = \left(\frac{\sigma_{Ga}}{\sigma_{In}} - F_{In}\right) c_{In} \tag{A2}$$

where $F_{In}$ is the fraction of In atoms that substitutes for Ga atoms in the alloy, and $\sigma_{Ga}$ and $\sigma_{In}$ are the scattering cross sections for Ga and In, respectively [38]. Assuming In replaces Ga at a 1:1 ratio in the alloy (for every 1 gallium atom lost, 1 indium atom is gained at the lattice site), then $F_{In} = 1$.

$$\sigma = \frac{Z^{\frac{4}{3}} \lambda^2 [1 + \frac{E_0}{m_0 c^2}]^2}{\pi [1 + (\frac{\beta}{\theta_0})^2]} \tag{A3}$$

Here, $E_0$ and $\lambda$ are the incident electron energy and wavelength, respectively, $m_0$ is the electron rest mass, $\beta$ is the angle of collection of the objective aperture (> 50 mrad for HAADF), and $\theta_0$ is the characteristic screening angle. Using $V = 200$ kV incident beam, the relativistically corrected electron wavelength is

$$\lambda = \sqrt{\frac{1.5}{V + 10^{-6}V^2}} = 2.5 \times 10^{-12} \text{ m} \tag{A4}$$

Using the bulk Bohr radius of $a_{InAs} = 34$ nm [39] and $a_{GaAs} = 11.6$ nm [40].

$$\theta_0 = \frac{\lambda Z^{1/3}}{2\pi a} \tag{A5}$$

Solving for the relativistic energy and scattering cross sections

$$\frac{E_0}{m_0 c^2} \approx 1.39 \tag{A6}$$

and

$$\sigma_{Ga} \approx 5.16314 \times 10^{-27} \text{ m}^2 \tag{A7}$$

$$\sigma_{In} \approx 1.49303 \times 10^{-27} \text{ m}^2 \tag{A8}$$

$$c_{In} = \frac{C}{\frac{\sigma_{Ga}}{\sigma_{In}} - F_{In}} \approx 0.31 \tag{A9}$$

Therefore, the QD core depicted in Fig. 6(a) has a peak In content of 31%.

APPENDIX B: APT DATA SET

Figs. 7(a) and (b) show all detection events that occurred during the APT experiment, plotted as a function of the mass-to-charge ratio. Each detection event corresponds to a location within the volume of the specimen, meaning that it has an associated position in 3D space. The histogram in Fig. 7(b) shows the cutoff that was used to minimize the statistical inclusion of $As_3^{2+}$ ions in the dataset, which has a similar mass-to-charge ratio to $^{113}In^+$.

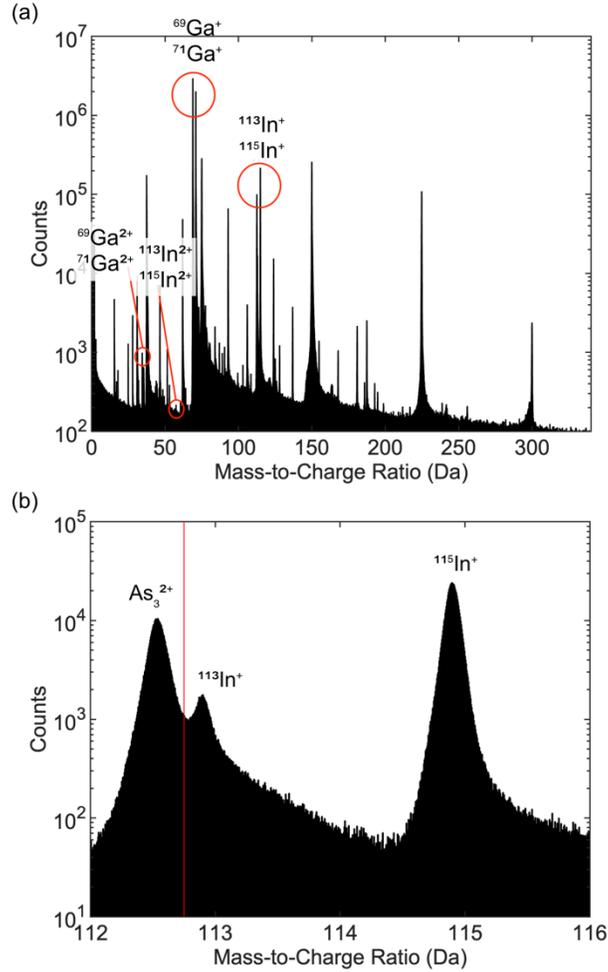

FIG. 7. (a) Histogram of total APT detection events. The labelled peaks have been used in the main text, and the remaining peaks correspond to different As and P ions. (b) APT counts in the region of $^{113}\text{In}^+$.

APPENDIX C: NEIGHBOR SEARCHING AND SPATIAL RESOLUTION

In Fig. 8, the maximum number of nearest neighbors required (to find an In or Ga atom) for data columns parallel to the growth axis is shown for the middle QD layer. On average, more voxels must be searched near the outer reaches of the APT sample, demonstrating the issue of signal compression associated with APT. Most columns near the center find an atom for every voxel within the single nearest neighbor search radius, whereas columns near the border of the data are more likely to contain voxels requiring double/triple neighbor searching.

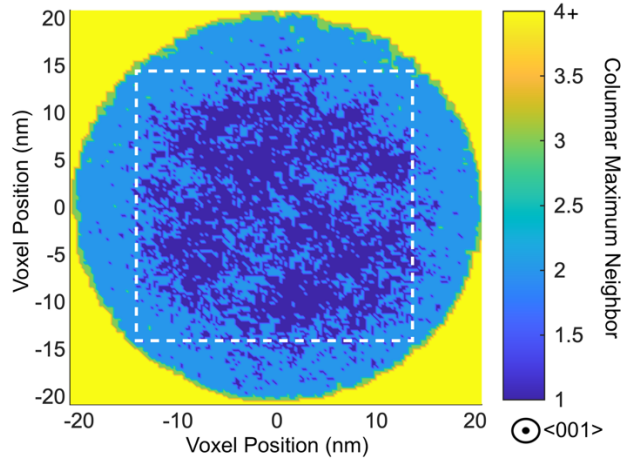

FIG. 8. 2-dimensional projection of the maximum neighbor search radius along each voxel column parallel to the growth axis. The dashed white box represents the region used to create the neighbor search probability distribution in Fig. 2. Due to the lower detection efficiency near the border, the spatial resolution outside the dashed white box is lower than the values reported in the main text.

The following discussion pertains to the determination of the spatial resolution of the modeled sample. A grid is decomposed into $N$ volume elements (with edge length $dx$, $dy$, and $dz$), using at most $K$ neighbors, and no constraints are imposed on minimum searching. The spatial resolution, $r_n$, for a voxel which uses $n$ neighbors is:

$$r_n(dx, dy, dz) = (2n+1)\sqrt{dx^2 + dy^2 + dz^2} \qquad (C1)$$

If $P_n$ is the probability of a voxel using $n$ neighbours, the mean, $\mu$, and variance, $\sigma^2$, of the spatial resolution are determined by the following expressions:

$$\mu = \sum_{n=0}^{K} P_n r_n \qquad (C2)$$

$$\sigma^2 = \sum_{n=0}^{K} P_n [r_n - \mu]^2 \qquad (C3)$$

After imposing a minimum one neighbor constraint on searching, the 0 and 1 neighbor search families are combined such that the probability is,

$$P_0 \cup P_1 = P_0 + P_1 - P_0 \cap P_1 = P_0 + P_1 \qquad (C4)$$

$P_0 \cap P_1 = 0$ since the two families are disjoint. The mean and variance become

$$\mu = \sum_{n=1}^{K} \begin{matrix} (P_0 + P_n)r_n & n = 1 \\ P_n r_n & n > 1 \end{matrix} \tag{C5}$$

$$\sigma^2 = \sum_{n=1}^{K} \begin{matrix} (P_0 + P_n)[r_n - \mu]^2 & n = 1 \\ P_n[r_n - \mu]^2 & n > 1 \end{matrix} \tag{C6}$$

Without neighbor searching, the minimum spatial resolution resulting in atoms in every voxel is 1.64 nm, shown by the black star on Fig. 9(a). Our neighbor search protocol enables the mean spatial resolution to be reduced below sub-nanometer scale, while ensuring no vacant voxels. Figs. 9(a) and (b) show that imposing a minimum constraint on searching drastically decreases the variance in the voxel size, but at the cost of a lower spatial resolution. Fig. 2 demonstrated decay of the $n = 1$ search family below a volume of ~$0.037a_0^3$, based on geometric considerations of the unit cell with grid decomposition, where $a_0$ is the lattice constant of GaAs. This is further visualized in Fig. 9(b), as the variance in voxel size increases rapidly for a minimum search of one nearest neighbor (min=1 family) below this voxel volume.
The spatial resolution after imposing a minimum 1 nearest neighbor search becomes better than the minimum obtainable without searching at $0.17a_0^3$. As the total number of voxels is inversely proportional to the voxel volume, choosing the largest voxel size without compromising the spatial resolution is desired. It was found that a voxel volume between $0.037a_0^3$ and $0.17a_0^3$ with a 1 neighbor minimum search constraint simultaneously yields good spatial resolution and variance. We note that in the limit of an infinitely small voxel size with unlimited neighbor searching allowed, the resolution of the reconstruction will approach the spatial resolution of the APT experiment itself.

The voxel dimensions in the main text are *dx, dy* = 0.25 nm and *dz* = 0.15 nm, resulting in a volume of ~$0.052a_0^3$. This corresponds to a spatial resolution of $1.1 \pm 0.1$ nm, shown by the orange star in Fig. 9(b). The *dz* edge length has been made to be 3/5 the *dx* (and *dy*) length, as the spatial resolution for the APT experiment in the z-direction (along the <001> growth axis) is ~3/5 the lateral resolution and the QDs have a low aspect ratio.

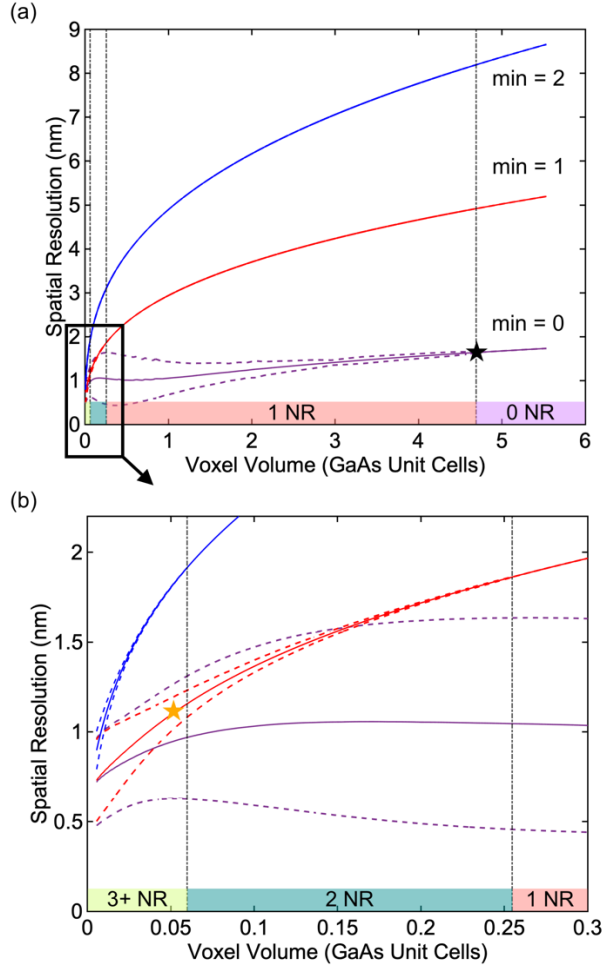

FIG. 9. Mean spatial resolution (solid) and single standard deviation (dashed) as a function of voxel volume for forced minimum neighbor searching of 0, 1 and 2 neighbors. The black vertical dot-and-dashed indicate when there is an increase in the number of neighbors required (NR) to completely fill all voxels with at least one group III ion. The black star in (a) indicates the lowest spatial resolution (voxel size) required to assign composition to the entire sample volume without neighbor searching. The orange star in (b) indicates the spatial resolution used in the present study.

APPENDIX D: $In_xGa_{1-x}As$ BAND PROFILE

If a voxel uses $n$ neighbours ($n \geq 1 \cap n \in \mathbb{Z}$) to find a single group III ion, there are a total of $(2n + 1)^3$ voxels used in the search. The indium fraction, $x$, assigned to the center voxel in the search is:

$$x = \frac{\sum_{i=1}^{(2n+1)^3} v_i(\text{In})}{\sum_{i=1}^{(2n+1)^3} v_i(\text{In}) + v_i(\text{Ga})} \tag{D1}$$

Where $v_i$ represents the total sum of indium or gallium counts within the $i^{th}$ voxel's bounds used in the search. For bulk $\text{In}_x\text{Ga}_{1-x}\text{As}$, the fundamental gap obeys the quadratic relationship [41],

$$E_g(x) = 1518 - 1580x + 475x^2 \quad \text{meV} \quad 0 \leq x \leq 1 \tag{D2}$$

The valence band edge as a function of the indium concentration, $x$, is [41]

$$E_v(x) = 231x - 58x^2 \quad \text{meV} \quad 0 \leq x \leq 1 \tag{D3}$$

Therefore, the hole barrier height maintaining $E_h(x = 1) = 0$ is

$$E_h(x) = E_v(x) - E_v(x = 1) = -58x^2 + 231x - 173 \text{ meV} \tag{D4}$$

Additionally, the electron barrier height maintaining $E_e(x = 1) = 0$ is

$$E_e(x) = E_g(x) + E_h(x) - E_g(x = 1) = 417x^2 - 1349x + 932 \text{ meV} \tag{D5}$$

This results in a band offset ratio of $E_e : E_h \approx 84.3 : 15.7$. The following discussion on strain is based on ref. [42], using linear elastic theory for $\text{In}_x\text{Ga}_{1-x}\text{As}$ single quantum wells and assumes no coupling between the light hole and heavy hole bands. The conduction band will shift to increase the bandgap due to hydrostatic pressure resulting from compressive biaxial strain within the plane of the QDs. Assuming the lattice to be biaxially strained to the in-plane GaAs lattice, the valence band will also shift to reduce the bandgap from both hydrostatic pressure and uniaxial tension [42], which would occur along the growth axis of the QDs. The relaxed lattice parameter changes as a function of concentration [42],

$$a_0(x) = 5.6536 + 0.4054x \text{ Å} \tag{D6}$$

The relative difference in lattice constant is:

$$\epsilon(x) = \frac{a_0(x) - a_0(0)}{a_0(0)} \tag{D7}$$

$E_H(x)$ and $E_S(x)$ are the hydrostatic and uniaxial strain energy given by [42],

$$E_H(x) = a(x)\big(2 - K(x)\big)\epsilon(x) \tag{D8}$$

$$E_S(x) = b(x)\big(1 + K(x)\big)\epsilon(x) \tag{D9}$$

Where $a(x)$ and $b(x)$ are the deformation potentials under hydrostatic and tetragonal perturbation. For an accurate representation of the deformation potentials, the bowing parameter has been used to interpolate the nonlinear relationship [43].

$$a(x) = -5920x - 8500(1-x) + 1470(x-x^2) \text{ meV} \tag{D10}$$
$$b(x) = -1680x - 2000(1-x) + 200(x-x^2) \text{ meV} \tag{D11}$$

Additionally,
$$K \equiv \frac{-2S_{12}}{S_{11} + S_{12}} \tag{D12}$$

where $S_{ij}$ is the elastic compliance tensor and can be related to the elastic stiffness tensor $C_{ij}$ through Hooke's law [44].

$$S_{11} = \frac{C_{11} + C_{12}}{(C_{11} - C_{12})(C_{11} + 2C_{12})} \tag{D13}$$

$$S_{12} = \frac{-C_{12}}{(C_{11} - C_{12})(C_{11} + 2C_{12})} \tag{D14}$$

Values for $C_{11}(x)$, $C_{12}(x)$, and $K(x)$ are given in [42]. The relative contribution of the hydrostatic strain to the conduction and valence bands is determined by the ratio of the pressure sensitivity of the spin orbit to the total shift in the bandgap under pressure, which has been experimentally determined for this lattice system [42].

$$\Delta Q_H = \frac{\delta(E_g + \Delta_0)/\delta P}{\delta(E_g)/\delta P} \approx 0.11 \tag{D15}$$

The shift in energy of the conduction and valence bands are [42]:
$$\Delta E_c(x) = -(1 - \Delta Q_H)E_H(x) \tag{D16}$$
$$\Delta E_v(x) = -\Delta Q_H E_H(x) + E_S(x) \tag{D17}$$

The potential barrier height in the Schrödinger equation can then be derived for each voxel in the strained system as

$$V_e(\text{x}) \equiv E_e(x) + \Delta E_c(x) + \frac{E_g(x=1)}{2} \tag{D18}$$

$$V_{hh}(\text{x}) \equiv E_h(x) - \Delta E_v(x) - \frac{E_g(x=1)}{2} \tag{D19}$$

The band offset ratio in the strained system changes to $V_e : V_{hh} \approx 59.3 : 40.7$. This change in offset ratio can be seen in Fig. 10(a) along with the band profile. The electron and heavy hole effective masses are given by [42],

$$m_e(x) = 0.067(1 - 0.426x)m_0 \tag{D20}$$
$$m_{hh}(x) = 0.34(1 + 0.117x)m_0 \tag{D21}$$

The static relative permittivity as a function of the indium concentration is [45]

$$\epsilon_r(x) = 12.4 + 2.15x \tag{D22}$$

The exciton Bohr radius is

$$a_{exciton}(x) = \frac{4\pi\epsilon_0 \epsilon_r(x) \hbar^2}{\mu(x) q^2} \tag{D23}$$

where $\mu$ is the reduced mass given by

$$\frac{1}{\mu(x)} = \frac{1}{m_e(x)} + \frac{1}{m_{hh}(x)} \tag{D24}$$

Substituting the appropriate values results in $11.7 \lessapprox a_{exciton} \lessapprox 22.1$ nm for $x \in [0,1]$, shown in Fig. 10(b).

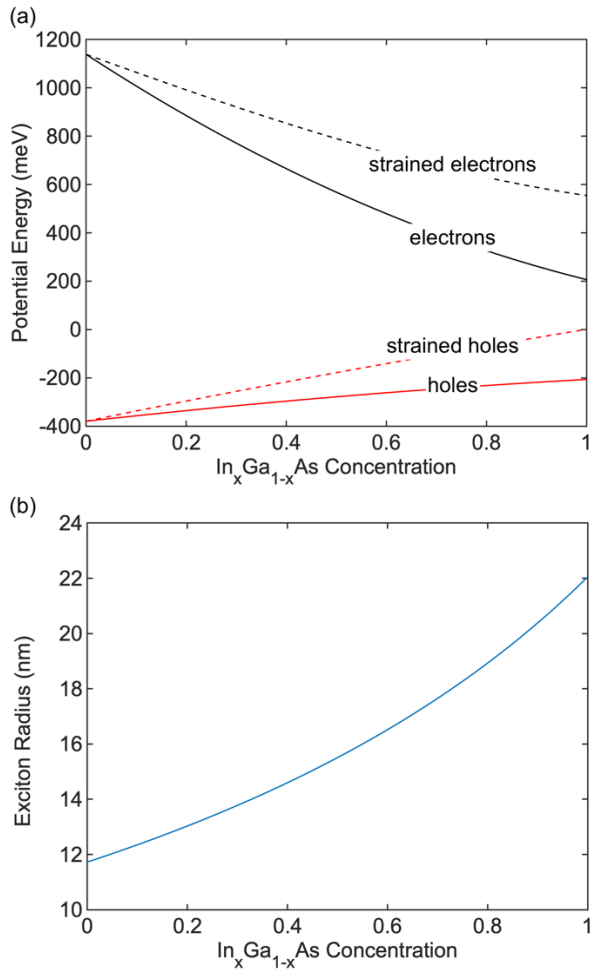

FIG. 10. (a) Calculated valence (red) and conduction (black) bands for the relaxed (solid) and strained (dashed) $In_xGa_{1-x}As$ system. (b) Calculated exciton radius for $In_xGa_{1-x}As$.

[1]   F. Grillot, D. Arsenijevic, D. Bimberg, and H. Huang, *Ultrafast and Nonlinear Dynamics of InAs/GaAs Semiconductor Quantum Dot Lasers*, in (2018), p. 22.
[2]   J. C. Norman, D. Jung, Z. Zhang, Y. Wan, S. Liu, C. Shang, R. W. Herrick, W. W. Chow, A. C. Gossard, and J. E. Bowers, *A Review of High-Performance Quantum Dot Lasers on Silicon*, IEEE J Quantum Electron **55**, 1 (2019).
[3]   K. Ludge and E. Scholl, *Quantum-Dot Lasers—Desynchronized Nonlinear Dynamics of Electrons and Holes*, IEEE J Quantum Electron **45**, 1396 (2009).
[4]   H. Liu, Q. Wang, J. Chen, K. Liu, and X. Ren, *MOCVD Growth and Characterization of Multi-Stacked InAs/GaAs Quantum Dots on Misoriented Si(100) Emitting near 1.3 Mm*, J Cryst Growth **455**, 168 (2016).
[5]   G. Burkard, D. Loss, and D. DiVincenzo, *Coupled Quantum Dots as Quantum Gates*, Phys Rev B **59**, 2070 (1998).
[6]   K. Eng et al., *Isotopically Enhanced Triple-Quantum-Dot Qubit*, Sci Adv **1**, e1500214 (2015).
[7]   M. Bayer, P. Hawrylak, K. Hinzer, S. Fafard, M. Korkusinski, Z. Wasilewski, O. Stern, and A. Forchel, *Coupling and Entangling of Quantum States in Quantum Dot Molecules*, Science **291**, 451 (2001).
[8]   C. Jennings, X. Ma, T. Wickramasinghe, M. Doty, M. Scheibner, E. Stinaff, and M. Ware, *Self-Assembled InAs/GaAs Coupled Quantum Dots for Photonic Quantum Technologies*, Adv Quantum Technol **3**, (2019).
[9]   S. Koley, J. Cui, Y. Panfil, and U. Banin, *Coupled Colloidal Quantum Dot Molecules*, Acc Chem Res (2021).
[10]  A. Mittelstädt, L. Greif née Hilse, S. Jagsch, and A. Schliwa, *Efficient Electronic Structure Calculations for Extended Systems of Coupled Quantum Dots Using a Linear Combination of Quantum Dot Orbitals Method*, Phys Rev B Condens Matter **103**, 115302 (2021).
[11]  I. Khanonkin, S. Bauer, V. Mikhelashvili, O. Eyal, M. Lorke, F. Jahnke, J. P. Reithmaier, and G. Eisenstein, *On the Principle Operation of Tunneling Injection Quantum Dot Lasers*, Prog Quantum Electron **81**, 100362 (2022).
[12]  M. Logar, S. Xu, S. Acharya, and F. B. Prinz, *Variation of Energy Density of States in Quantum Dot Arrays Due to Interparticle Electronic Coupling*, Nano Lett **15**, 1855 (2015).
[13]  D. Wigger et al., *Controlled Coherent Coupling in a Quantum Dot Molecule Revealed by Ultrafast Four-Wave Mixing Spectroscopy*, ACS Photonics **10**, 1504 (2023).
[14]  L. Mancini et al., *Three-Dimensional Nanoscale Study of Al Segregation and Quantum Dot Formation in GaAs/AlGaAs Core-Shell Nanowires*, Appl Phys Lett **10547**, (2014).
[15]  C. Greenhill, A. S. Chang, E. S. Zech, S. Clark, G. Balakrishnan, and R. S. Goldman, *Influence of Quantum Dot Morphology on the Optical Properties of GaSb/GaAs Multilayers*, Appl Phys Lett **116**, 252107 (2020).
[16]  E. Di Russo et al., *Super-Resolution Optical Spectroscopy of Nanoscale Emitters within a Photonic Atom Probe*, Nano Lett **20**, 8733 (2020).
[17]  L. Mancini, F. Moyon, D. Hernàndez-Maldonado, I. Blum, J. Houard, W. Lefebvre, F. Vurpillot, A. Das, E. Monroy, and L. Rigutti, *Carrier Localization in GaN/AlN Quantum Dots As Revealed by Three-Dimensional Multimicroscopy*, Nano Lett **17**, 4261 (2017).
[18]  I. Dimkou, D. R. Enrico, P. Dalapati, J. Houard, N. Rochat, D. Cooper, E. Bellet-Amalric, A. Grenier, E. Monroy, and L. Rigutti, *InGaN Quantum Dots Studied by Correlative*

[18] *Microscopy Techniques for Enhanced Light-Emitting Diodes*, ACS Appl Nano Mater **3**, (2020).
[19] N. Jeon, B. Loitsch, S. Morkoetter, G. Abstreiter, J. Finley, H. Krenner, G. Koblmüller, and L. Lauhon, *Alloy Fluctuations Act as Quantum Dot-like Emitters in GaAs-AlGaAs Core-Shell Nanowires*, ACS Nano **9**, (2015).
[20] G. Mantashyan, P. Mantashyan, and D. Hayrapetyan, *Modeling of Quantum Dots with the Finite Element Method*, Computation **11**, 5 (2023).
[21] R. V. N. Melnik and M. Willatzen, *Bandstructures of Conical Quantum Dots with Wetting Layers*, Nanotechnology **15**, 1 (2004).
[22] W. C. Yek, G. Gopir, and A. P. Othman, *Calculation of Electronic Properties of InAs/GaAs Cubic, Spherical and Pyramidal Quantum Dots with Finite Difference Method*, in *Advanced Materials Research*, Vol. 501 (2012), pp. 347–351.
[23] L. Gong, Y. C. Shu, J. J. Xu, Q. S. Zhu, and Z. G. Wang, *Numerical Analysis on Quantum Dots-in-a-Well Structures by Finite Difference Method*, Superlattices Microstruct **60**, 311 (2013).
[24] R. Dash and S. Jena, *Finite Element Analysis of the Effect of Wetting Layer on the Electronic Eigenstates of InP/InGaP Pyramidal Quantum Dots Solar Cell*, in *Materials Today: Proceedings*, Vol. 39 (Elsevier Ltd, 2019), pp. 2015–2021.
[25] A. Fakkahi, S. Dahiya, M. Jaouane, A. Ed-Dahmouny, R. Arraoui, A. Sali, M. N. Murshed, H. Azmi, and N. Zeiri, *Finite Element Analysis of Multilayered Spherical Quantum Dots: Effects of Layer Dimensions, Alloy Composition, and Relaxation Time on the Linear and Nonlinear Optical Properties*, Physica B Condens Matter **690**, (2024).
[26] J.-P. Leburton, J. Destiné, P. Matagne, J.-P. Leburton, J. Destine, and G. Cantraine, Modeling of the Electronic Properties of Vertical Quantum Dots by the Finite Element Method, 2000.
[27] M. Jaouane, R. Arraoui, A. Ed-Dahmouny, A. Fakkahi, K. El-Bakkari, H. Azmi, and A. Sali, *Finite Element Method Simulation of Electronic and Optical Properties in Multi-InAs/GaAs Quantum Dots*, Eur Phys J Plus **139**, (2024).
[28] M. E. Mora-Ramos et al., *Electronic Structure of Vertically Coupled Quantum Dot-Ring Heterostructures under Applied Electromagnetic Probes. A Finite-Element Approach*, Sci Rep **11**, (2021).
[29] H. Kim and J. Kim, *Extended Dimensionality of the Density of States in InGaAs Coupled Quantum Dot-Ring Structures as Evidenced by Radiative Decay Times*, Appl Surf Sci **624**, 156932 (2023).
[30] B. Gault, A. Chiaramonti, O. Cojocaru-Mirédin, P. Stender, R. Dubosq, C. Freysoldt, S. Makineni, T. Li, M. Moody, and J. Cairney, *Atom Probe Tomography*, Nature Reviews Methods Primers **1**, 51 (2021).
[31] B. Gault, M. P. Moody, J. M. Cairney, and S. P. Ringer, *Analysis Techniques for Atom Probe Tomography*, in *Atom Probe Microscopy*, edited by B. Gault, M. P. Moody, J. M. Cairney, and S. P. Ringer (Springer New York, New York, NY, 2012), pp. 213–297.
[32] T. Borrely, T.-Y. Huang, Y.-C. Yang, R. Goldman, and A. Quivy, *On the Importance of Atom Probe Tomography for the Development of New Nanoscale Devices*, in (2022), pp. 1–4.
[33] C. Schimpf, M. Reindl, F. Basso Basset, K. D. Jöns, R. Trotta, and A. Rastelli, *Quantum Dots as Potential Sources of Strongly Entangled Photons: Perspectives and Challenges for Applications in Quantum Networks*, Appl Phys Lett **118**, 100502 (2021).

[34] D. Ferreira, J. Sousa, R. Maronesi, J. Bettini, M. Schiavon, A. Teixeira, and A. Silva, *Size-Dependent Bandgap and Particle Size Distribution of Colloidal Semiconductor Nanocrystals*, J Chem Phys **147**, (2017).

[35] L. E. Brus, *Electron–Electron and Electron-hole Interactions in Small Semiconductor Crystallites: The Size Dependence of the Lowest Excited Electronic State*, J Chem Phys **80**, 4403 (1984).

[36] Y. Kayanuma, *Quantum-Size Effects of Interacting Electrons and Holes in Semiconductor Microcrystals with Spherical Shape*, Phys Rev B **38**, 9797 (1988).

[37] K. Nishi, T. Kageyama, M. Yamaguchi, Y. Maeda, K. Takemasa, T. Yamamoto, M. Sugawara, and Y. Arakawa, *Molecular Beam Epitaxial Growths of High-Optical-Gain InAs Quantum Dots on GaAs for Long-Wavelength Emission*, J Cryst Growth **378**, 459 (2013).

[38] D. B. Williams and C. B. Carter, *Amplitude Contrast*, in *Transmission Electron Microscopy: A Textbook for Materials Science*, edited by D. B. Williams and C. B. Carter (Springer US, Boston, MA, 2009), pp. 371–388.

[39] H. Fang, H. A. Bechtel, E. Plis, M. C. Martin, S. Krishna, E. Yablonovitch, and A. Javey, *Quantum of Optical Absorption in Two-Dimensional Semiconductors*, Proc Natl Acad Sci U S A **110**, 11688 (2013).

[40] P. Harrison and A. Valavanis, *Quantum Wells, Wires and Dots: Theoretical and Computational Physics of Semiconductor Nanostructures* (2016).

[41] A. Schliwa, M. Winkelnkemper, and D. Bimberg, *Impact of Size, Shape, and Composition on Piezoelectric Effects and Electronic Properties of In(Ga)As/GaAs Quantum Dots*, Phys Rev B **76**, 205324 (2007).

[42] D. J. Arent, K. Deneffe, C. Van Hoof, J. De Boeck, and G. Borghs, *Strain Effects and Band Offsets in GaAs/InGaAs Strained Layered Quantum Structures*, J Appl Phys **66**, 1739 (1989).

[43] P. A. Khomyakov, M. Luisier, and A. Schenk, *Compositional Bowing of Band Energies and Their Deformation Potentials in Strained InGaAs Ternary Alloys: A First-Principles Study*, Appl Phys Lett **107**, 062104 (2015).

[44] S. Adachi, *Mechanical, Elastic, and Lattice Vibrational Properties*, in *Physical Properties of III-V Semiconductor Compounds* (1992), pp. 17–47.

[45] E. H. Li, *Material Parameters of InGaAsP and InAlGaAS Systems for Use in Quantum Well Structures at Low and Room Temperatures*, Physica E Low Dimens Syst Nanostruct **5**, 215 (2000).